\documentclass[sigconf]{acmart}

\AtBeginDocument{%
  \providecommand\BibTeX{{%
    \normalfont B\kern-0.5em{\scshape i\kern-0.25em b}\kern-0.8em\TeX}}}

\setcopyright{acmcopyright}
\copyrightyear{2022}
\acmYear{2022}

\acmConference[WWW '22]{Proceedings of the ACM Web Conference 2022}{April 25--29, 2022}{Virtual Event, Lyon, France}
\acmBooktitle{Proceedings of the ACM Web Conference 2022 (WWW '22), April 25--29, 2022, Virtual Event, Lyon, France}
\acmPrice{15.00}
\acmDOI{10.1145/3485447.3512283}
\acmISBN{978-1-4503-9096-5/22/04}

\usepackage[inline]{enumitem}

\usepackage{xspace}
\usepackage{xcolor}
\usepackage{soul}
\usepackage{caption}
\usepackage{subcaption}

\newcommand\ws[1]{\texttt{#1}}

\newcommand\camera[1]{#1}

\def\eg{\emph{e.g.}\xspace}
\def\etc{\emph{etc.}\xspace}
\def\ie{\emph{i.e.}\xspace}

\def\vs{\emph{vs.}\xspace}
\def\cf{\emph{cf.}\xspace}

\begin{document}

\title{``Way back then'': A Data-driven View of 25+ years of Web Evolution}



\author{Vibhor Agarwal}
\affiliation{%
  \institution{Department of Computer Science\\University of Surrey}
  \city{Guildford}
  \state{Surrey}
  \country{United Kingdom}
}
\email{v.agarwal@surrey.ac.uk}

\author{Nishanth Sastry}
\affiliation{%
 \institution{Department of Computer Science\\University of Surrey}
 \city{Guildford}
 \state{Surrey}
 \country{United Kingdom}
}
\email{n.sastry@surrey.ac.uk}

\begin{abstract}
Since the inception of the first web page three decades back, the Web has evolved considerably, from static HTML pages in the beginning to the dynamic web pages of today, from mainly the text-based pages of the 1990s to today's multimedia rich pages, \etc{}. Although much of this is known anecdotally, to our knowledge, there is no quantitative documentation of the extent and timing of these changes.
This paper attempts to address this gap in the literature by looking at the top 100 Alexa websites for over 25 years from the Internet Archive or the ``Wayback Machine'', \ws{archive.org}. We study the changes in popularity, from Geocities and Yahoo! in the mid-to-late 1990s to the likes of Google, Facebook, and Tiktok of today. We also look at different categories of websites and their popularity over the years and find evidence for the decline in popularity of news and education-related websites, which have been replaced by streaming media and social networking sites. We explore the emergence and relative prevalence of different MIME-types (text \vs{} image \vs{} video \vs{} javascript and json) and study whether the use of text on the Internet is declining. 

\end{abstract}




\maketitle

\section{Introduction}
\label{introduction}

Since the first webpage was put up by Sir Tim Berners-Lee at CERN 30 years ago\footnote{\url{http://info.cern.ch/hypertext/WWW/TheProject.html}, last accessed 9 Dec 2021.}, the World Wide Web has expanded dramatically year on year, and now contains at least 1,200 petabytes of data~\cite{starry2019howbig}. It is estimated that the size of the Web follows Moore's Law~\cite{mooreslaw} and doubles in every few months.
Despite this wealth of knowledge and information now available at our fingertips, the Web can be an ephemeral experience as thousands of URLs become unreachable and thousands of new URLs are created every minute. Therefore, capturing the history of the Web is an important and challenging exercise.

Anecdotally speaking, the Web and its technologies as well as the dominant and popular platforms or websites have changed considerably since its inception. Although some of the fundamental underlying technologies such as HTML and HTTP have remained relatively stable despite newer versions, the visual appearance of the top websites as well as which websites are popular can and have changed significantly over the years. For example, GeoCities, an extremely popular web hosting site in the 1990s, is no longer present, but new sites such as TikTok are gaining traction rapidly.

In this paper, we are interested in enhancing our quantitative understanding of the above kinds of change. \camera{To examine this, we focus on the Top 100 websites in the Web, as determined by \texttt{alexa.com}, a popular source for ranking and categorising websites (all ranks as determined in Nov 2021). We also examine Google Trends data, which allows us to compare the relative popularity of two search terms or websites over time. Finally,  we utilise data from the Wayback machine or archive.org, which takes periodic snapshots of much of the Web to get a historic view of these websites from up to 25 years back.}

Developing a broad and over-arching quantitative historical understanding of the evolution of the Web requires us to move beyond individual websites and to study \textit{large collections} \camera{of} websites and web resources, which are identified by Uniform Resource Locators (URLs)\footnote{\camera{Datasets and code are available at \url{https://github.com/socsys/wayback-web-history}.}}. Therefore, although our datasets have rich data about individual websites and even individual URLs, we look at them \textit{collectively}, asking questions about how users spend their time on different categories of websites as well as how the popularity of individual websites within a category (such as social media), or the relative popularity of different categories of websites (\eg{} news \vs{} social media), have changed over time. We also group together URLs by their MIME types\footnote{ MIME stands for Multimedia Internet Mail Extension and MIME types were developed in the context of email. IANA has changed the terminology and MIME types are now called media types~\cite{mediatypes} as their use has extended well beyond emails. However, it appears that the usage ``MIME type'' is more common, \eg{} one of our primary data sources, \ws{archive.org}, uses MIME rather than media types. Therefore this paper also uses MIME type as it may be more familiar to the readers.}, which have historically been used to identify different kinds of multimedia content, and thus provide a convenient lens to study whether and to what extent the Web has moved away from its initial mainly (hyper)text-based origins.

Our analysis reveals a number of interesting trends. For example, we show evidence for decreased popularity of news and education websites, which have been replaced by streaming media and social networking. Using Google Trends, we are able to chart and compare the rise or decline of competing websites, such as different social media platforms, which may all be competing for a similar user base. We can also show, using the \ws{archive.org} data, how the number of URLs in each website has been increasing steadily over time, and how the Web is increasingly turning to images, videos and other sophisticated forms of content. Despite these changes, we still find that text dominates in terms of the \textit{number} of web resources (\ie{} number of Uniform Resource Locators or URLs) available on most website categories. The time correlations that can be observed suggest that some of the changes, such as increased usage of JavaScript Object Notation (JSON) objects or JavaScript URLs may be related to the rise in popularity of web programming frameworks such as AJAX and Representational State Transfer (REST).

The rest of this paper is organised as follows. Section~\ref{sec:background} discusses related work including other attempts to take a historical look at the evolution of the Web. Section~\ref{sec:datasets} describes the datasets we use and how we collect it. Section~\ref{sec:dynamics} then explores how the popularity of different websites and different website categories has changed in the last 25+ years. Section~\ref{sec:mimetypes} studies the MIME types of different URLs and the evolution of multimedia and other forms of content on the Web. Section~\ref{sec:discussion} concludes with a discussion of our findings, limitations of our approach and directions for future work.

\section{Background and Related work}
\label{sec:background}

 One of the earliest accounts we could find is a history of hypertext in the Web, from one of the founding figures of the Web conference, Robert Cailliau\cite{cailliau1999hypertext}. Maddux~\cite{maddux1997world} focuses on educational website design. \cite{foot2010object} introduces the idea of a historiography of the Web. \camera{\cite{aya2006building} discusses how to build a research library for the history of the Web.}

Within the last decade, considerable scholarly interest has emerged in the Web's history. A few studies by Brugger and others~\cite{brugger2017web, brugger2011web, ben2018internet, balbi2018history, bru2018archived} have pioneered more recent approaches to creating a history of the Web but have mostly focused on specific perspectives rather than broad quantitative understanding: \cite{brugger2011web} focuses on the consequences of \camera{the} archiving process for the Internet scholar. \cite{brugger2017web} studies history of media and government, size of web domains, and cultural and political histories. \cite{ben2018internet} analyses the socio-technical epistemic processes behind the construction of historical facts by the Internet Archive, Wayback Machine. It uncovers specific epistemic processes through a case-study on the archiving of the North Korean Web. \cite{jackson2012formats} explores the UK Web history. It looks at image, HTML, and PDF resources, and shows how the usage of different formats, versions, and software implementations has changed over time.

Yet, to our knowledge there is no comprehensive review and the history of the Web remains relatively understudied. Initial attempts that approach include Br{\"u}gger et al.~\cite{brugger2018sage}, who  provides a multifaceted understanding of the \camera{web history}. Goggin et al.~\cite{goggin2017routledge} brings together the diverse Internet histories that have evolved in different regions, languages, and social contexts across the globe. \cite{arora2014leisure} uses spatial metaphors to talk about web history.



In popular perception, the history of the Web is divided into multiple versions~\cite{shivalingaiah2008comparative}: Web 1.0 refers to the first stage of the World Wide Web evolution with static web pages based on basic HTML templates. Web 2.0 refers to worldwide websites which highlight user-generated content, usability, and interoperability for end users. Web technologies  used in Web 2.0 development include AJAX and JavaScript frameworks for dynamic content. Web 3.0 refers to the evolution of web utilization and interaction which includes altering the Web into a database and adding semantics. \cite{allen2013web} studies what we learn from major ``versions'' of the Web. The paper explores Web 2.0 as the marker of a discourse about the nature and purpose of the Internet in the recent past. It focuses on how Web 2.0 introduced to our thinking about the Internet a discourse of versions. Such a discourse enables the telling of a ‘history’ of the Internet which involves a complex interweaving of past, present and future, as represented by the additional versions which the introduction of Web 2.0 enabled.

Our quantitative approach relies heavily on web archives. Web Archiving is any form of deliberate and purposive preserving of web material~\cite{brugger2011web}. Web Archiving began in 1996~\cite{brugger2018sage} with a stated purpose of preserving the digital culture of the world. The Internet Archive was created in 1996 as a non-profit organization to preserve historical collections that exist in digital format. The Web collection is based on web crawls performed by Alexa Internet, a for-profit organization and crawls are done based on where the links point and where the users go. The main archiving strategy is the \textit{snapshot} approach, where all the web material that the web crawler encounters is archived periodically~\cite{brugger2011web}. The frequency of archival depends upon the rate at which a website is updated, \eg{}, Google's snapshot is taken from hundreds to thousands of times per day\footnote{\url{https://web.archive.org/web/*/www.google.com}, last accessed 9 Dec 2021.}. \camera{Till date, archive.org has collected more than two decades of Web history, through regular snapshots of websites, preserved, and made them accessible.} \cite{rogers2017doing} suggests to use Internet archive for Web history as it provides websites homepages as they were in the past.

We also rely on Google Trends, a freely accessible tool that allows users to interact with Internet search data and may provide deep insights into population behavior and other social phenomena~\cite{nuti2014use}. It gives data based on the search queries on Google Search. \cite{choi2012predicting} uses Google trends data to predict important tasks such as unemployment claims, consumer confidence, automobile sales, etc. \cite{nuti2014use} studies the importance of Google trends data in healthcare research. Stephens-Davidowitz~\cite{stephens2013essays} pioneered the use of Google trends data. In the book ``Everybody lies''~\cite{stephens2017everybody}, he talks about Google search as a ``truth serum'' as it is not intended to present or project a certain persona to everyone (so searches may reveal information such as personal illnesses which may have a stigma attached to them). In this study, we also use Google trends data as an indicator of the popularity of websites over time. The interest in a particular search keyword over time gives the popularity of that keyword/topic.

There are other tools and datasets which could potentially be used, but are not considered in this paper. For example, the use of historical crawls of the web such as the Common Crawl\footnote{\url{https://commoncrawl.org}, last accessed 26 Jan 2022.}. \cite{mirtaheri2014brief} is a brief history of web crawlers. \cite{ben2016does} considers reconstructing history of deleted websites through a case study of the \texttt{.yu} Top Level Domain (which was removed after yugoslavia ceased to be a separate country).

Future historians using Web Archive data should be aware of attacks such as possible rewriting of web history~\cite{lerner2017rewriting}. We assume in this study that the archival data we look at have not been tampered with. Archive data can also be biased. For example, \cite{thelwall2004fair} explores the (un)fairness of archive.org based on country balance.
\cite{stock2017web} presents other angles, such as a historical overview of client side insecurities in the web browser.

The two main gaps in the literature which we hope to fill are \begin{enumerate*}
\item a \textit{quantitative} understanding of web history
\item a study of how the Web has evolved  over a long period of 25+ years.
\end{enumerate*} \camera{This is our long-term research agenda, and this paper represents a first attempt towards this goal.}

\section{Methodology}
\label{sec:datasets}
In this section, we describe our methodology for curating a selection of websites  and crawling the historical  data of these websites.

\subsection{Websites List Curation}
\label{sec:curation}
Alexa is an Internet company which provides web analysis including website traffic statistics, website comparisons, and website audience. Alexa's measurement panel includes a diverse set of 25,000 browser extensions and plug-ins used by millions of Internet users~\cite{greg2017alexa}. We collect the top 100 websites globally from Alexa.com\footnote{\url{https://www.alexa.com/topsites}, last accessed 8 Dec 2021.} during November 2021, according to the Alexa traffic rank. The Alexa rank data is derived from the traffic data provided by the users in Alexa's Global Data Panel over a rolling 3 months period and is based on the browsing pattern of people~\cite{determine-alexa-ranks}. We also get information -- ``Daily Time on Site'' which is an estimated daily time spent per visitor on the website, and ``Daily Pageviews per Visitor'' which is an estimated daily unique pageviews on the website, from Alexa.com.
For each website, we also obtain the category of the website, using fortiguard.com\footnote{\url{https://fortiguard.com}, last accessed 3 Dec 2021.}. In total, we get 24 distinct website categories within these top 100 Alexa websites.

\subsection{Crawling Historical Websites Data}
\label{sec:historicaldata}
To get historical website data, we use the ``Wayback Machine'', also known as Internet Archive --- archive.org\footnote{\url{https://web.archive.org}, last accessed on 3 Dec 2021.}. We crawl  historical information about each of the 100 websites of \S\ref{sec:curation}. \camera{We are able to go  as far back as  1998 and collect data up to November 2021} from archive.org using selenium\footnote{\url{https://www.selenium.dev}, last accessed 3 Dec 2021.}. Selenium automates web browsing and the Selenium WebDriver drives a browser natively, as a user would. For example, to get google.com related data, we crawl the summary view from archive.org\footnote{\url{https://web.archive.org/details/google.com}, last accessed 3 Dec 2021.}. We programmatically change the range of years in the web form on archive.org to 1 year (starting from 1998 all the way to 2021) and get information related to different MIME-types (text, application, javascript, images, videos, etc.) such as the number of captured URLs, and new URLs for each MIME-type per year. In total, we crawl 2400 (=24*100) Internet archive pages for these top 100 Alexa websites across 24 years (from 1998 to 2021).

We also use Google Trends to get data related to the popularity of search topics or keywords  worldwide~\cite{google-trends-data} for each year. Google Trends data is available from 2004 onwards, so we obtain information from 2004--2021. For the popular search engine and social networking websites, we crawl Google Trends data for their search interest over time as it can be taken as an indicator of the popularity of a website. The value returned by Google Trends is termed as  ``Interest Over Time'' and represents the volume of searches relative to the highest point on the chart. Hence, an ``Interest Over Time'' value of 100 stands for the highest number of searches within a set of values returned. As an example, if we compare two search terms ``foo'' (which receives 1k, 5K and 2k searches respectively over a three month duration under study) and ``bar'' (which was searched 5K, 10K and 8K searches in the same three month duration), then 10K is taken as ``100'', and foo will have ``Interest Over Time'' values of 10, 50 and 20, in comparison with bar, which will have values of 50, 80 and 100. Because popularity is computed on a relative basis for a given set of search terms and normalised to a value of 100, the popularity of different websites can be compared within a single Google Trends comparison (in Figure~\ref{fig:google-trends-lineplots}), \textit{but comparison is not possible across the sub-figures}. \camera{We call this measure of popularity as the ``Google Trends popularity'' to distinguish it from other possible measures of popularity (\eg{} number of monthly active users or number of app installs of a website)}.

\camera{We also use data of the number of visits during each year from 1996--2019 to the top 10 websites across the world that year, compiled from media and benchmark reports by YouTube user ``Data is Beautiful''~\cite{youtube-top-sites}. We cross-check that this data is accurate by comparing with Wikipedia, which maintains similar data from 2012 onwards, based on Alexa historical data~\cite{wikipedia-top-websites}. Although this page lists the \textit{current} top sites according to Alexa, Wikipedia's edit history function reveals previous Alexa rankings. It so happens that this page has been maintained by Wikipedia contributors since 2012, allowing us to verify  the data collected by ``Data is Beautiful'' from 2012--19. We also use Wikipedia to extend the data beyond 2019, to include the number of visits to the top 10 websites in 2020 and 2021.}

\section{Dynamics of Web Popularity}
\label{sec:dynamics}
In this section, we analyse how the Web has evolved over the years, by first looking at the relative popularity  of different website categories currently and then examining how this has changed over the years.

\subsection{Popularity of different website categories}

\begin{figure*}
\centering
    \begin{subfigure}{0.30\textwidth}
      \centering
      \includegraphics[width=\linewidth]{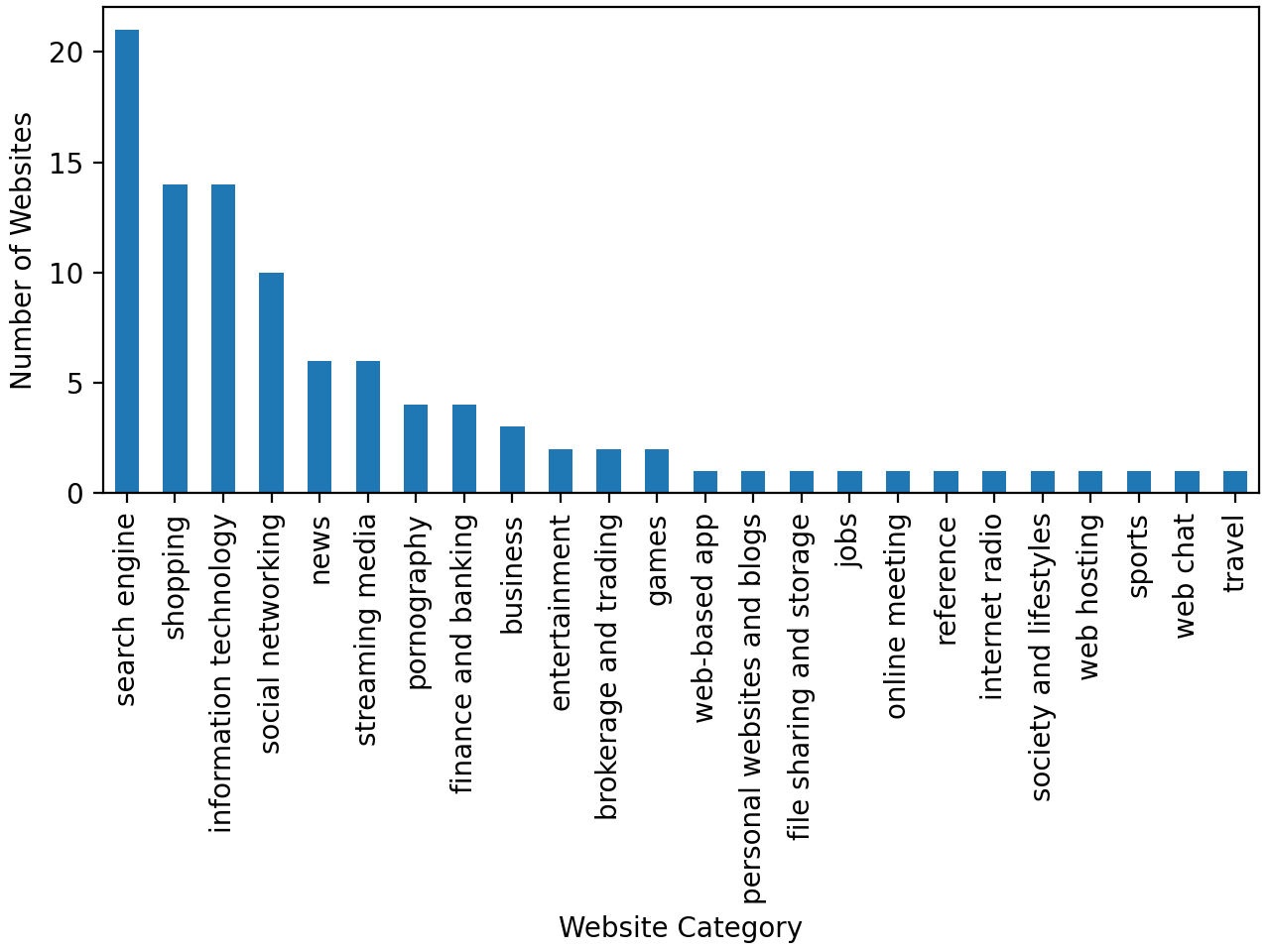}
      \caption{Categories of top 100 Alexa websites}
      \label{fig:websites-category}
    \end{subfigure}
    \begin{subfigure}{0.30\textwidth}
      \centering
      \includegraphics[width=\linewidth]{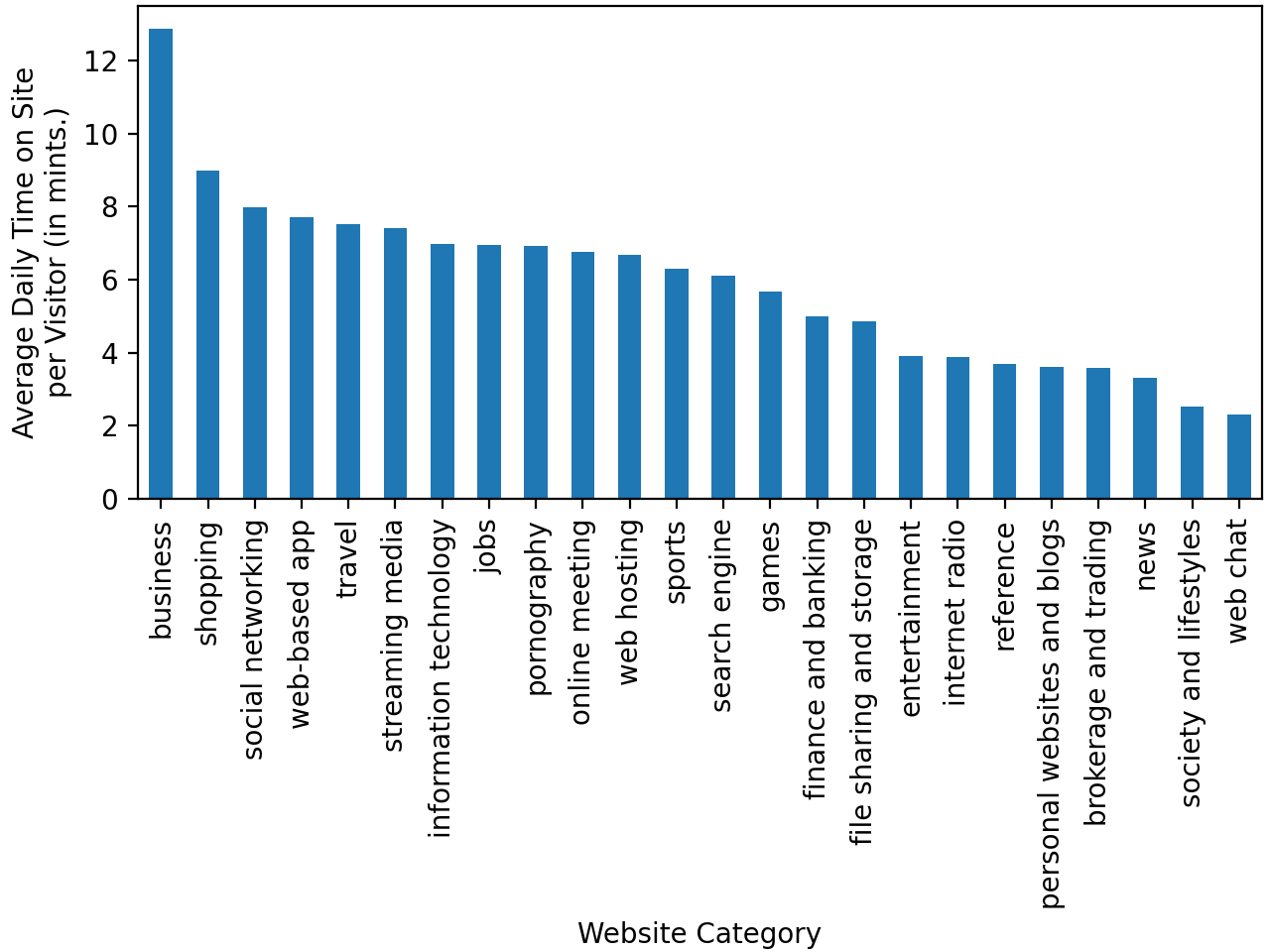}
      \caption{Average daily time (in minutes) spent on the website per visitor.}
      \label{fig:avg-daily-time}
    \end{subfigure}
    \begin{subfigure}{0.30\textwidth}
      \centering
      \includegraphics[width=\linewidth]{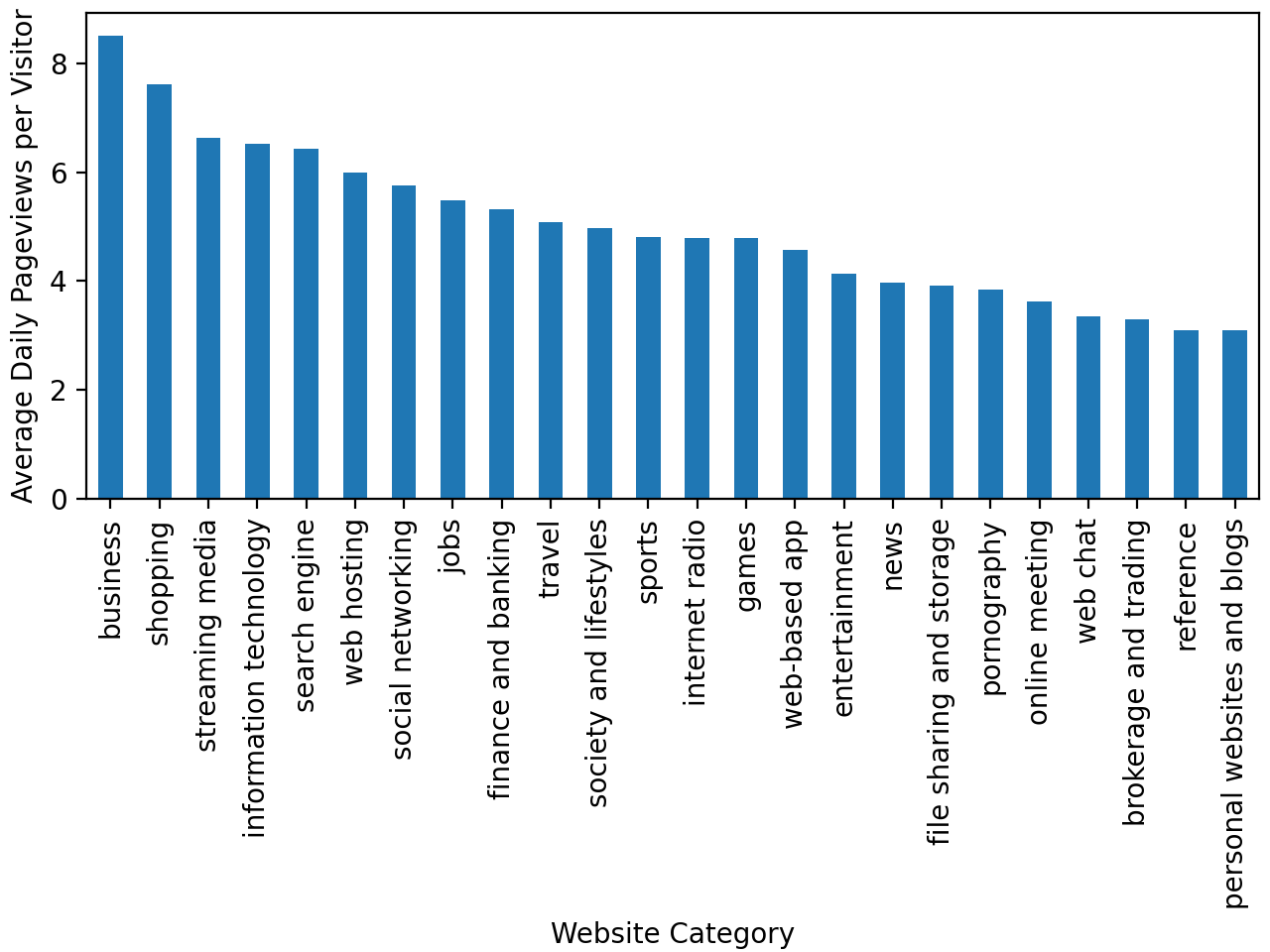}
      \caption{Average daily pageviews on the website per visitor.}
      \label{fig:avg-pageviews}
    \end{subfigure}
\caption{Different bar plots with website categories on x-axis, using the websites traffic data collected from \ws{Alexa.com}.}
\label{fig:alexa-barplots}
\end{figure*}

We first look at the website categories represented in the top 100 Alexa websites and the collective popularity of the set of websites from each category. Figure~\ref{fig:websites-category} shows the categories represented in the Alexa top 100. The highest number of websites (21) belongs to the search engine category, followed by shopping (14), information technology (14), and social networking (10). Out of 21 search engine domains, 7 domains are Google variants adapted for different countries, such as \ws{google.com} (USA and default for worldwide), \ws{google.co.in} (Google India), or \ws{google.co.jp} (Google Japan). Other search engines include Yandex (Russia) or Baidu (China) which are popular in specific countries. Interestingly, four adult and pornographic websites appear in the top 100, making them as numerous as the finance and banking category (also four websites). \camera{From a long time, pornographic and adult content has been driving media choices --- \eg{} Betamax \vs{} VHS, but also driving the sales of early pamphlets, news papers, \etc{}~\cite{coopersmith1998pornography}.}

The \textit{number} of websites in a category can be different from the total \textit{time} spent by users in a particular category of websites. Figure~\ref{fig:avg-daily-time} shows the average daily time spent on the website per visitor for each category \camera{over a 3-month period from Sep to Nov 2021}. Surprisingly, visitors spend the most time on business-related websites daily, followed by shopping websites. In our list, business websites such as \ws{Alibaba.com} are at the top, likely because of the additional services being offered by Alibaba and other websites. Social networking websites are at the third position, with visitors spending an average 8 minutes daily. The social network Facebook is at the top with 18 minutes as the daily time spent per visitor, followed by Twitter (12.5 minutes), and LinkedIn (11.25 minutes). These values may be indicative of how Alexa.com computes the time on each website: As it is based on toolbars installed in the browsers of its panel of users, we conjecture that the sojourn time on any website is based on new websites being loaded, and therefore may be missing pages being updated dynamically using AJAX and other techniques. Thus, the above timings might represent a lower bound on the actual times spent on each website.

Figure~\ref{fig:avg-pageviews} shows a different measure of how frequently each category of website is used, by measuring average daily page views per visitor \camera{over a 3-month period from Sep to Nov 2021} for each of the website categories. Again, business websites have the highest (9) unique daily page views per visitor. However, other categories such as social media, which typically rely on AJAX to dynamically load more content to the same web page on scrolling, have fewer distinct \textit{number of pages} recorded and therefore slip in the rankings.

\subsection{Changes in Popularity}
\label{subsec:popularity-changes}


\begin{figure*}
\centering
    \begin{subfigure}{0.30\textwidth}
      \centering
      \includegraphics[width=\linewidth]{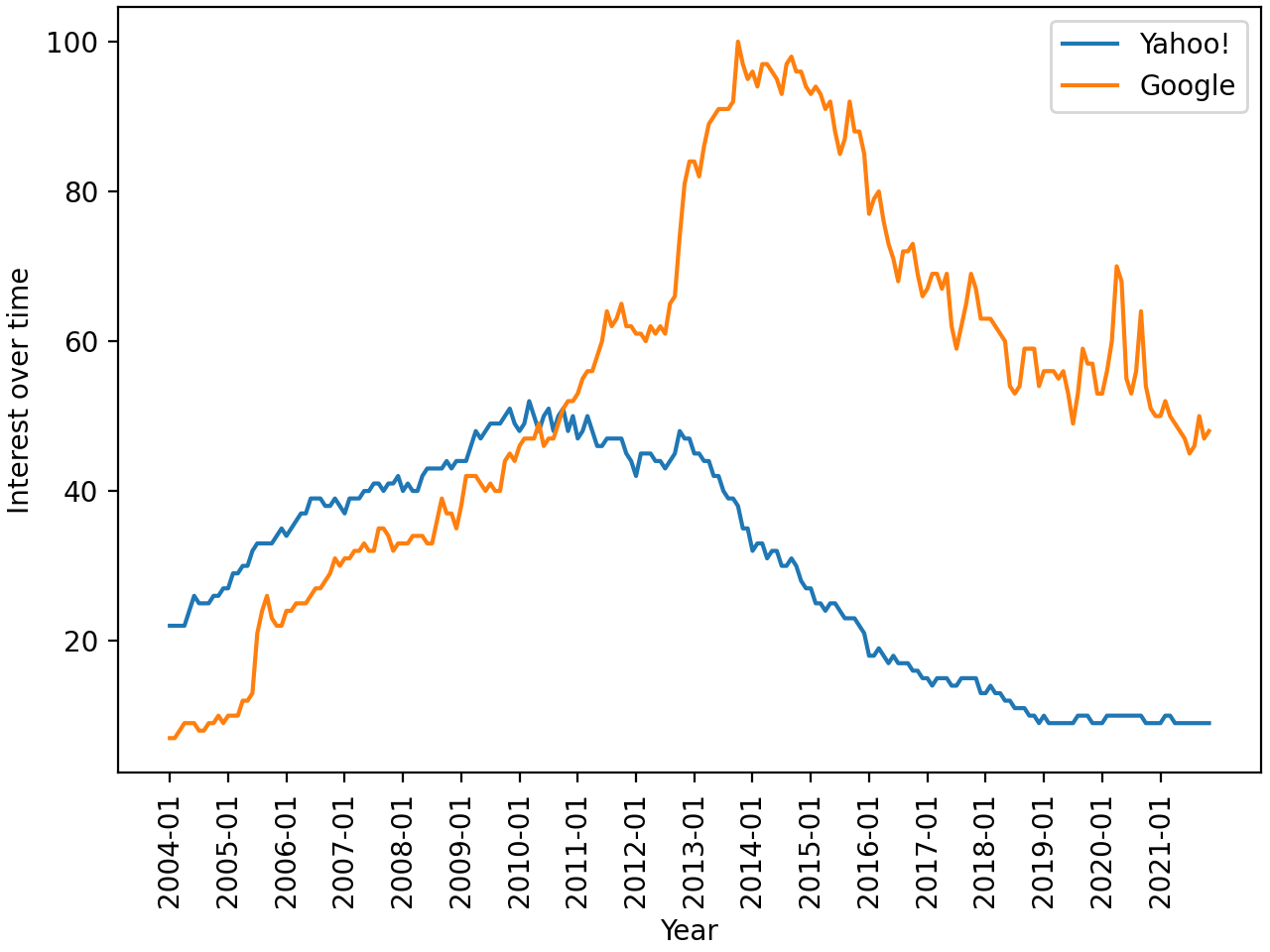}
      \caption{Yahoo! vs. Google.}
      \label{fig:yahoo-google}
    \end{subfigure}
    \begin{subfigure}{0.30\textwidth}
      \centering
      \includegraphics[width=\linewidth]{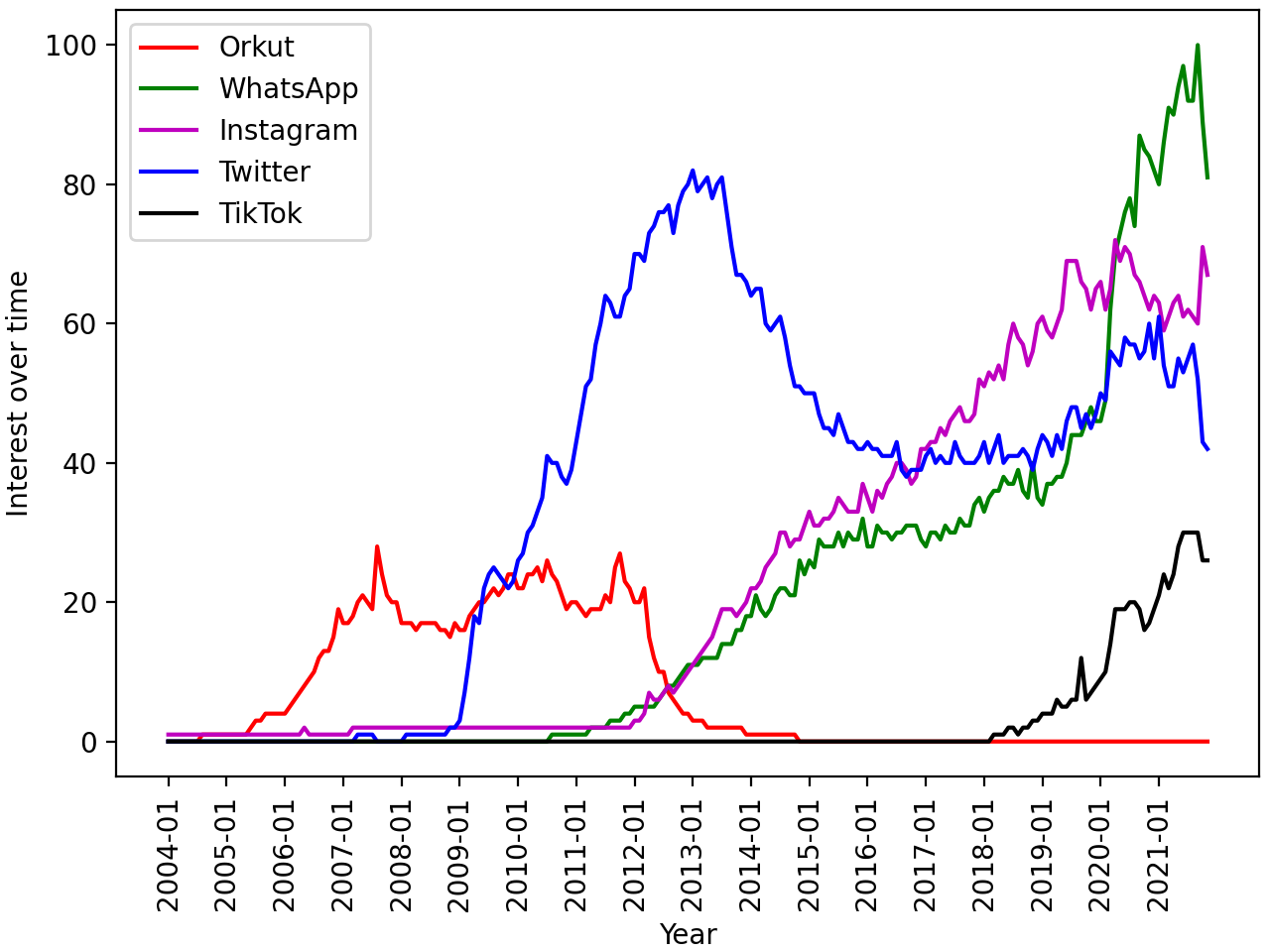}
      \caption{Social Networking websites.}
      \label{fig:social-network}
    \end{subfigure}
    \begin{subfigure}{0.30\textwidth}
      \centering
      \includegraphics[width=\linewidth]{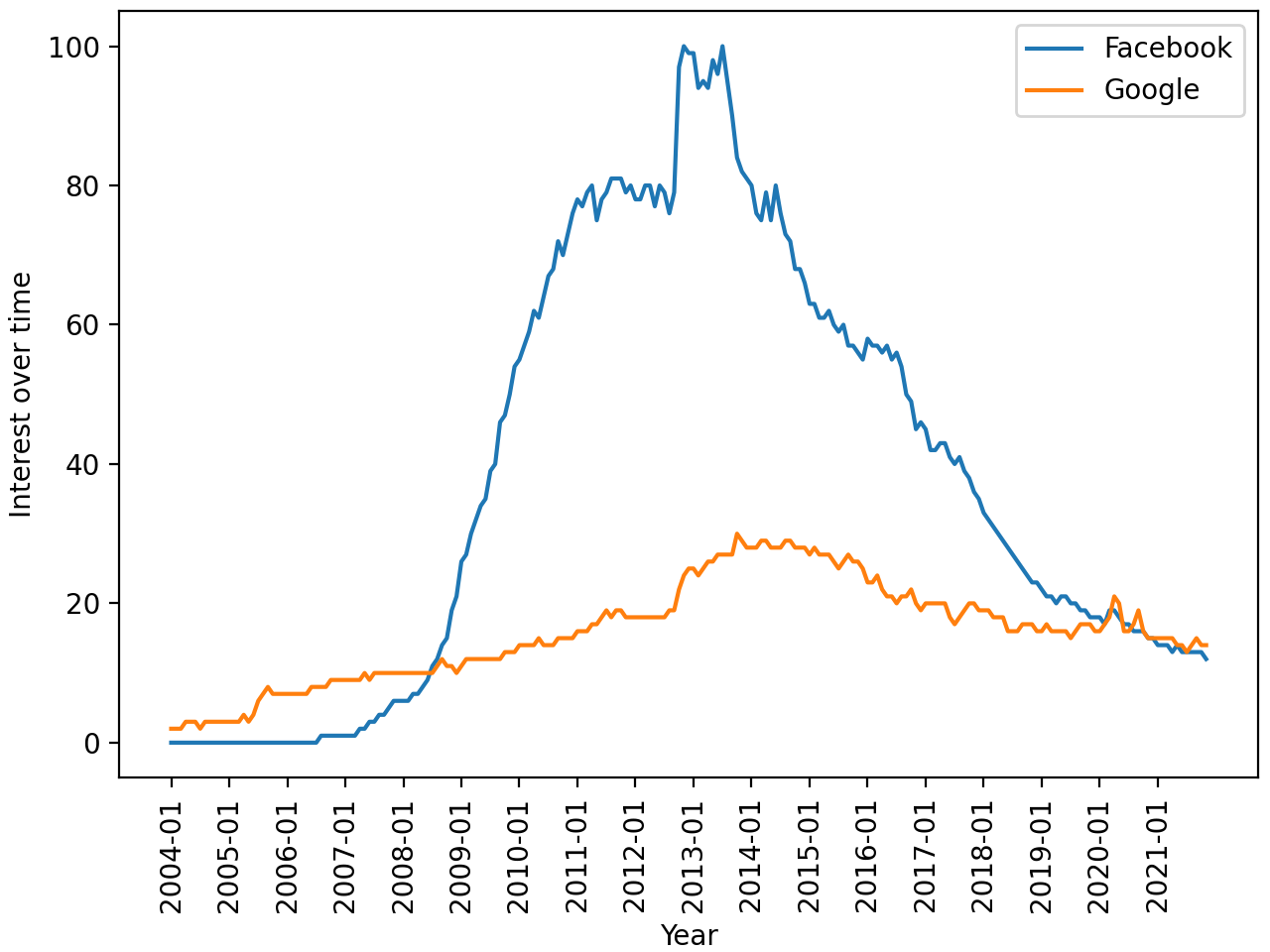}
      \caption{Facebook vs. Google.}
      \label{fig:fb-google}
    \end{subfigure}
\caption{\camera{Google Trends Popularity} of different websites based on the search interest over time.}
\label{fig:google-trends-lineplots}
\end{figure*}

The previous results show the relative popularity of different categories of websites \textit{today}. However, the popularity of a website can change dramatically over time (\eg{} once dominant search engines such as \ws{Altavista}, or turnkey web page creation and hosting services such as \ws{GeoCities}  are no longer around). To illustrate this dynamic and capture changes in popularity, we turn to Google Trends, which can provide (normalised) data about the relative popularity of two or more search keywords or topics over time (\cf{} \S\ref{sec:historicaldata}).

As a first case study, Figure~\ref{fig:yahoo-google} shows the relative number of searches for Google and Yahoo!, two dominant search engines. Google Trends captures every search, including search terms entered by people in the URL bar of their browsers, which are commonly designed as a combined URL bar and a search bar. \camera{Google Chrome, when first released in 2008, introduced this combined URL and search bar feature called Omnibox~\cite{jaber2014omnibox}. Other major browsers such as Mozilla Firefox introduced this feature in 2014~\cite{firefox2014omnibox}}. Thus, if a user enters the name of a website rather than its full URL, this results in a search in the default search engine (which may likely be Google), and therefore ends up being captured in Google Trends as one data point. Note that Google Trends only captures the searches on \ws{google.com}. It is therefore interesting to note that despite this obvious bias, there were \textit{more searches for Yahoo!} until nearly 2011 (\ie{} even users whose default search engine was Google, as configured either manually or as setup automatically upon installation, were preferring to search for Yahoo!, likely because of the additional services being offered on the Yahoo! Portal). We also observe that the relative number of searches for Google on Google.com (likely a result of users typing in ``google'' into their browsers' URL/Search bars) has been in decline since 2015. Some believe that this may be a result of increasing shift towards app-based access\footnote{https://www.businessinsider.com/the-number-of-people-using-search-engines-is-in-decline-2015-9, last accessed 21 Jan 2022.}.


As a second case study, we look at the category of  social networking websites. Social networks tend to be a ``winner take all'' market~\cite{prakash2012winner} as users need to be on the same platform as their friends. Therefore, a handful of websites end up capturing large proportions of users. Figure~\ref{fig:social-network} explores how the \camera{Google Trends popularity} of major social networks changed over time. In 2004, the earliest date for Google Trends data, \ws{Orkut} and \ws{MySpace} were the most widely used social networking websites, \camera{as evident from their relative popularity in Google Trends when compared with other prominent social networking websites}. Only around 2008 were they beaten by \ws{Facebook}, and later on, by \ws{Twitter}, based on the number of Google Searches made. Since 2010, the micro-blogging website Twitter became very popular and attained its peak in 2013. Around 2012, WhatsApp and Instagram came into existence and started to capture some of the Google Searches. By 2017, Instagram beat Twitter \camera{in Google Trends popularity}. In 2019, WhatsApp beat Instagram \camera{in Google Trends popularity} and became the most popular instant messaging application in the world, \camera{according to Google Trends data}. In 2021, WhatsApp and Instagram are the two most popular social networking websites, \camera{according to Google Trends data}. TikTok, a short video streaming application introduced in 2017, is rapidly gaining traction but has still not yet reached the same volume of searches \camera{on Google} as the other sites. In contrast, Facebook is not shown as its search volume is much higher, making it difficult to visualise the changes in popularity of the other sites. Figure~\ref{fig:fb-google} compares the popularity of the two of the most popular websites, \ws{Facebook} and \ws{Google}. It can be seen that in 2008 Facebook surpassed Google in \camera{Google Trends popularity} and reached its peak in 2013-14. After 2018 Cambridge Analytica Scandal, Facebook's reputation has been affected~\cite{peruzzi2018news} and its \camera{Google Trends popularity} has deteriorated rapidly.


\begin{figure}[h]
  \centering
  \includegraphics[width=\linewidth]{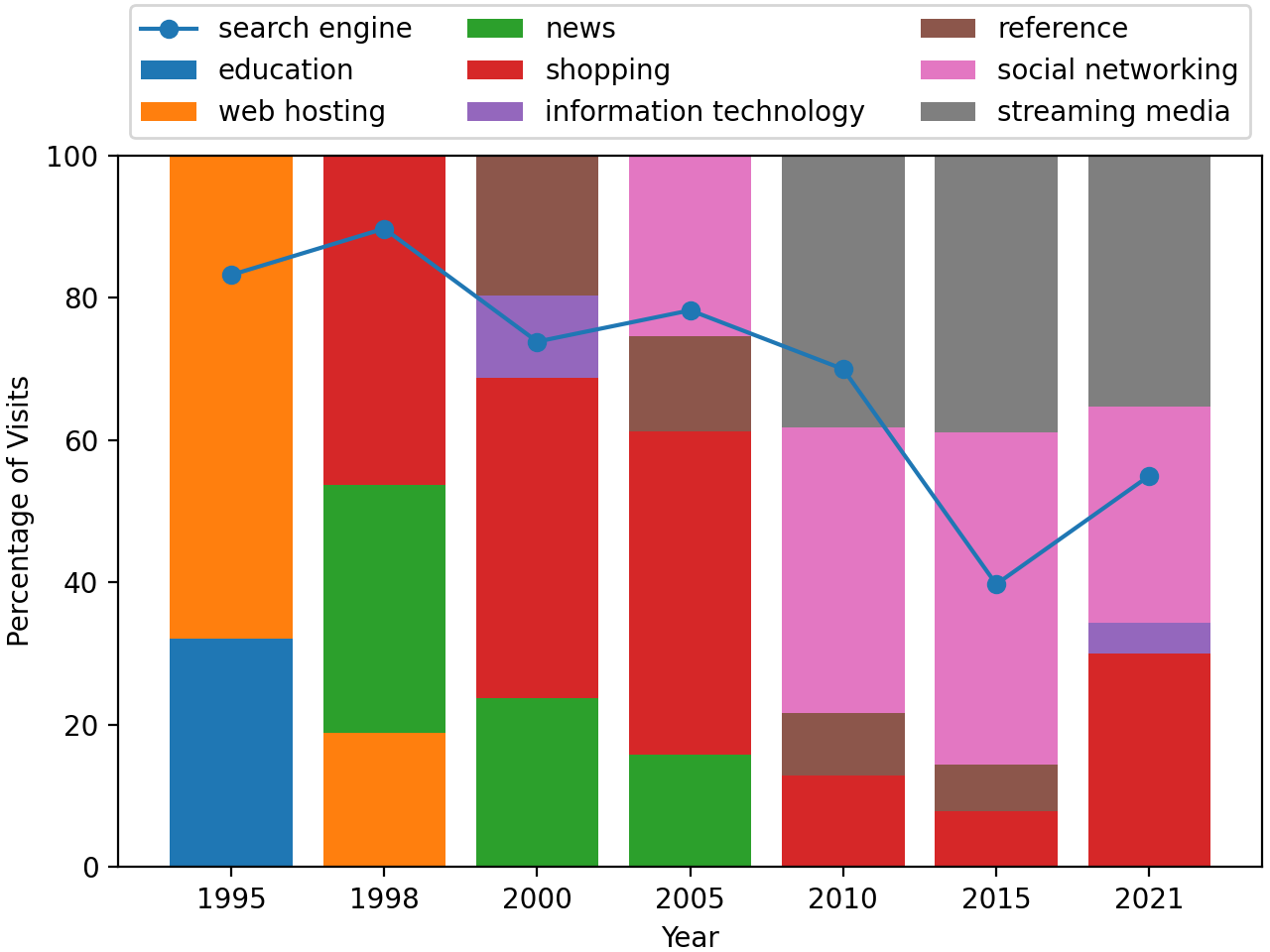}
  \caption{Percentage of visits to different website categories.}
  \label{fig:yearly-top-10-new}
\end{figure}

Given this constant flux in the popularity of different websites, we wished to investigate whether users have been always interested to the same extent in the same categories of websites and different websites end up being dominant in each category at different time points (\eg{} replacing MySpace and Orkut with Facebook in the social networking category without changing the proportion of visits to social networking sites), or whether users' interests have fundamentally changed over time.

To measure the relative popularities of different categories of websites, we use data of the number of visits during each year from 1996--2019 to the top 10 websites across the world that year, compiled from media and benchmark reports by YouTube user ``Data is Beautiful''~\cite{youtube-top-sites} \camera{(\cf{} description in  Section~\ref{sec:historicaldata})}.


Figure~\ref{fig:yearly-top-10-new} shows the percentage of visits captured by the top websites of different categories in each year. Note that the category of search engines capture a large proportion of visits, and so this category is shown separately as a line. In the initial years (1995 to 1998), search engine websites (Yahoo!, Altavista \etc{}) secured more than $80\%$ of the visits. Their visits percentage slowly dropped as new categories of websites became popular.

The stacked bar chart of Figure~\ref{fig:yearly-top-10-new} shows the fraction of visits to the categories of websites other than the dominating category of search (normalised to 100\% after removing search). This reveals a number of interesting trends: Web hosting sites such as GeoCities, a highly popular category in 1995--98 time frame, practically disappears from the top 10 by 2000. Similarly, education disappears after the initial years of popularity of the Web. News shows a decreasing trend and disappears from the top 10 after 2010. These categories are replaced by social network sites, which emerge around 2005, and streaming media, which becomes popular around 2010. We also note the new rise in popularity of online shopping sites in 2021, presumably a side-effect of the COVID-19 pandemic.


\section{Increase in Website Complexity}
\label{sec:mimetypes}

\begin{figure*}
\centering
    \begin{subfigure}{0.45\textwidth}
      \centering
      \includegraphics[width=\linewidth]{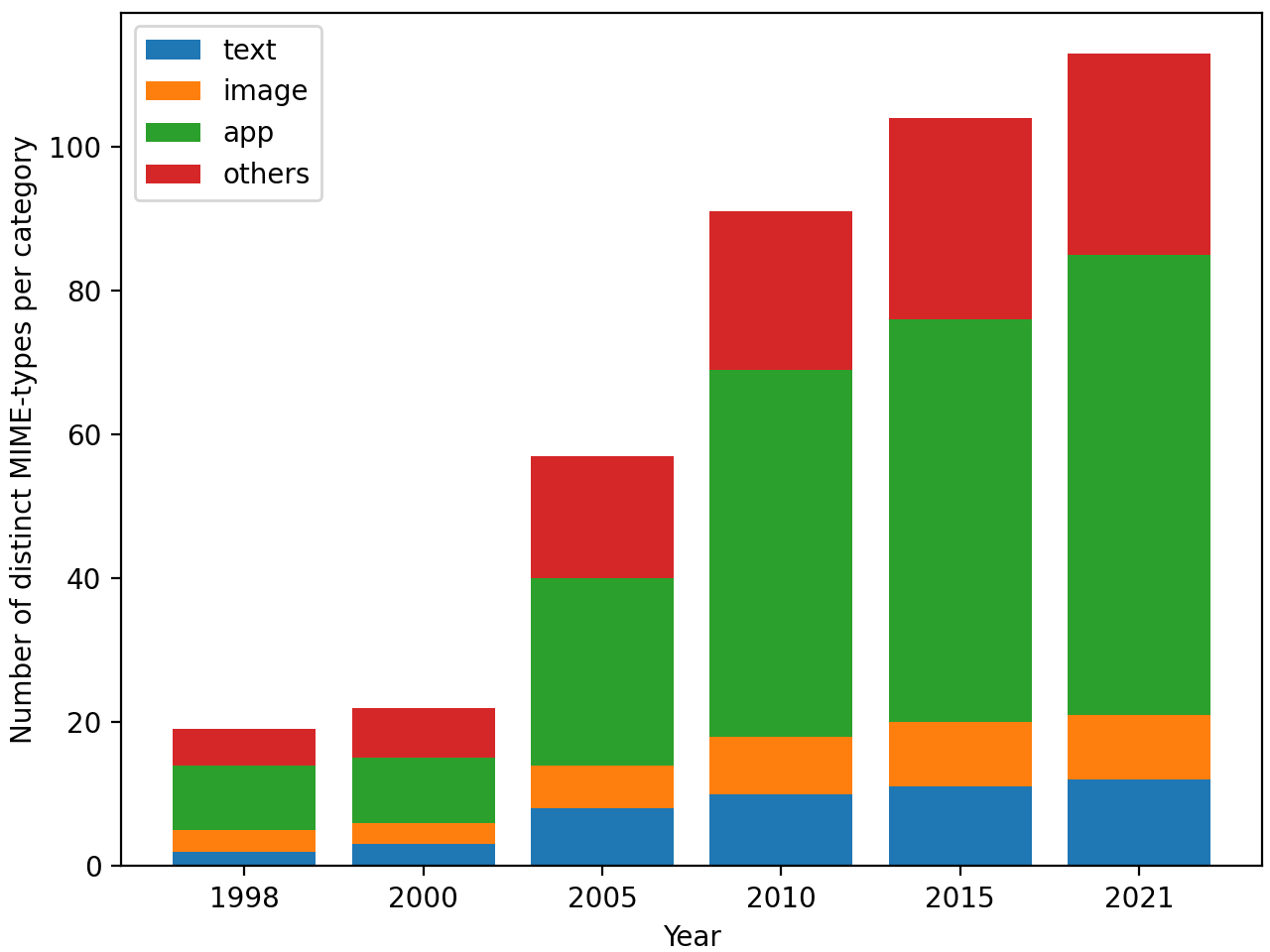}
        \caption{Number of distinct types for each MIME category per year.}
        \label{fig:num-mimetypes}
    \end{subfigure}
    \begin{subfigure}{0.45\textwidth}
      \centering
      \includegraphics[width=\linewidth]{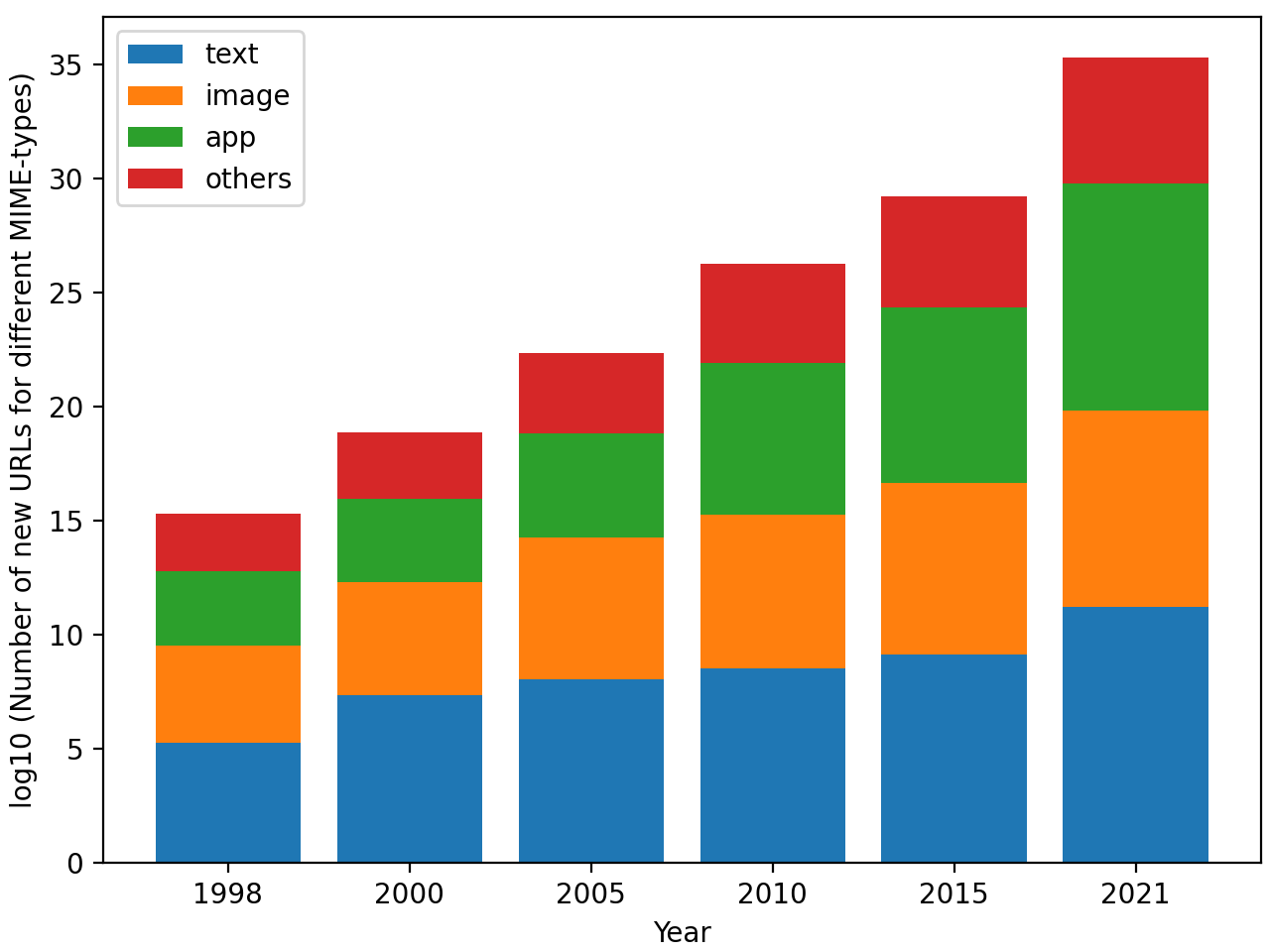}
      \caption{Number of new URLs for different MIME-types for all the top 100 websites across the years, on log10 scale.}
      \label{fig:yearly-mime}
    \end{subfigure}
\caption{Number of distinct MIME-types and URLs with these MIME-types.}
\label{fig:mime-stacked-barplots}
\end{figure*}


\begin{figure*}
\centering
    \begin{subfigure}{0.45\textwidth}
      \centering
      \includegraphics[width=\linewidth]{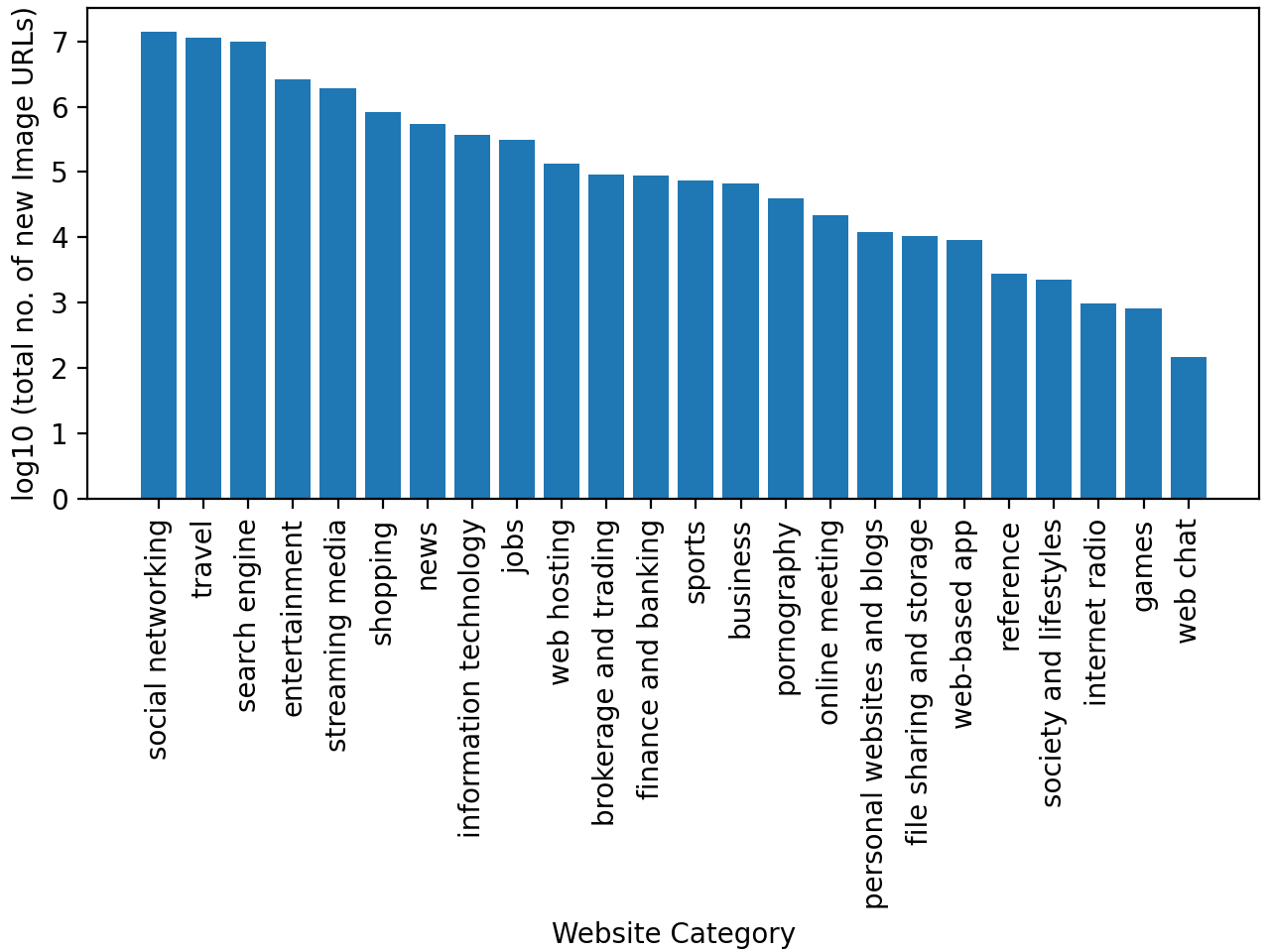}
      \caption{The total number of new Image URLs for each website category, normalized by the number of websites per category and sorted in descending order.}
      \label{fig:new-img-urls}
    \end{subfigure}
    \begin{subfigure}{0.45\textwidth}
      \centering
      \includegraphics[width=\linewidth]{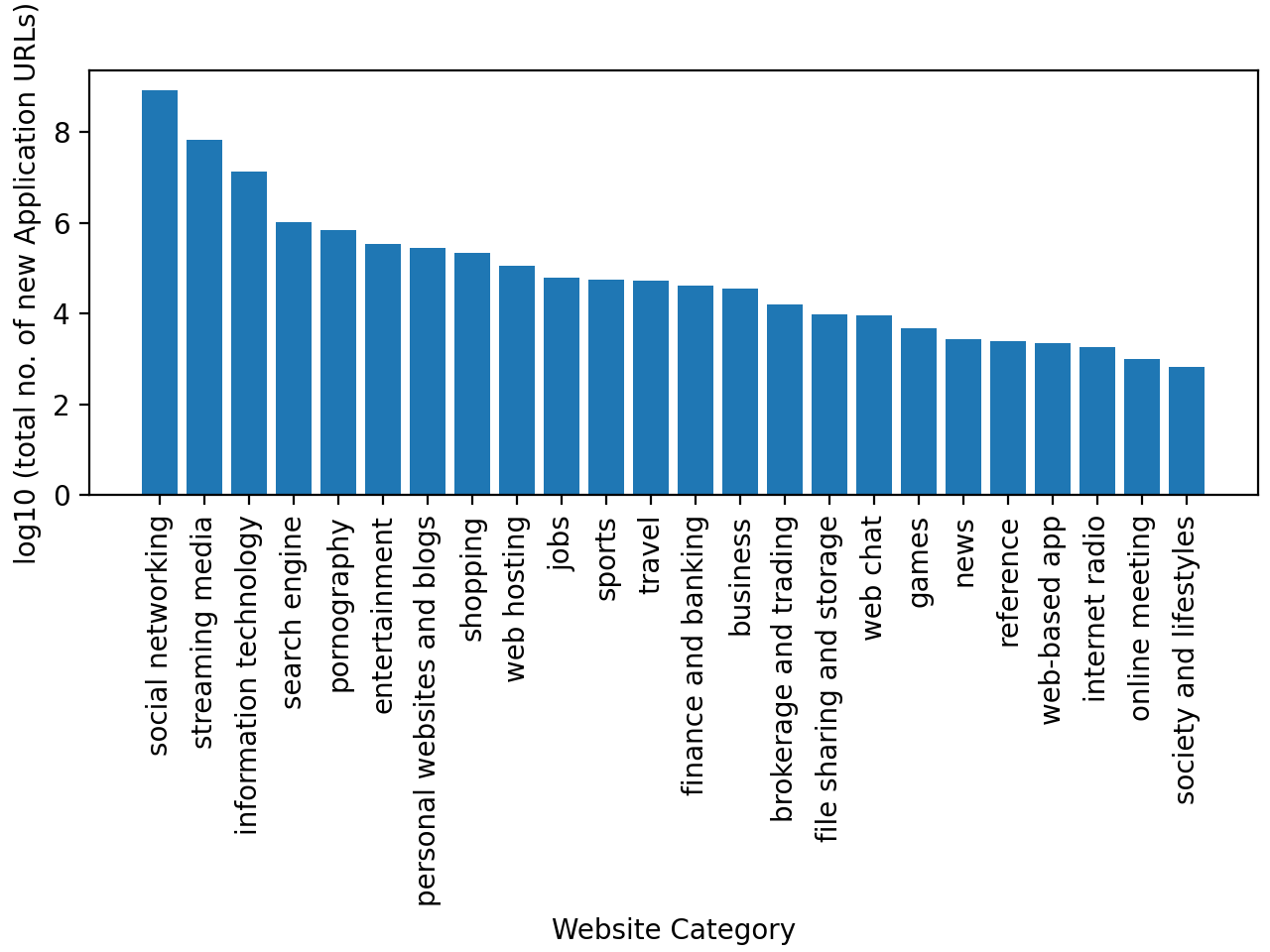}
      \caption{The total number of new Application URLs for each website category, normalized by the number of websites per category and sorted in descending order.}
      \label{fig:new-js-urls}
    \end{subfigure}
\caption{Number of URLs on top websites and their MIME-types.}
\label{fig:archive-barplots}
\end{figure*}

\begin{figure*}
\centering
    \begin{subfigure}{0.45\textwidth}
      \centering
      \includegraphics[width=\linewidth]{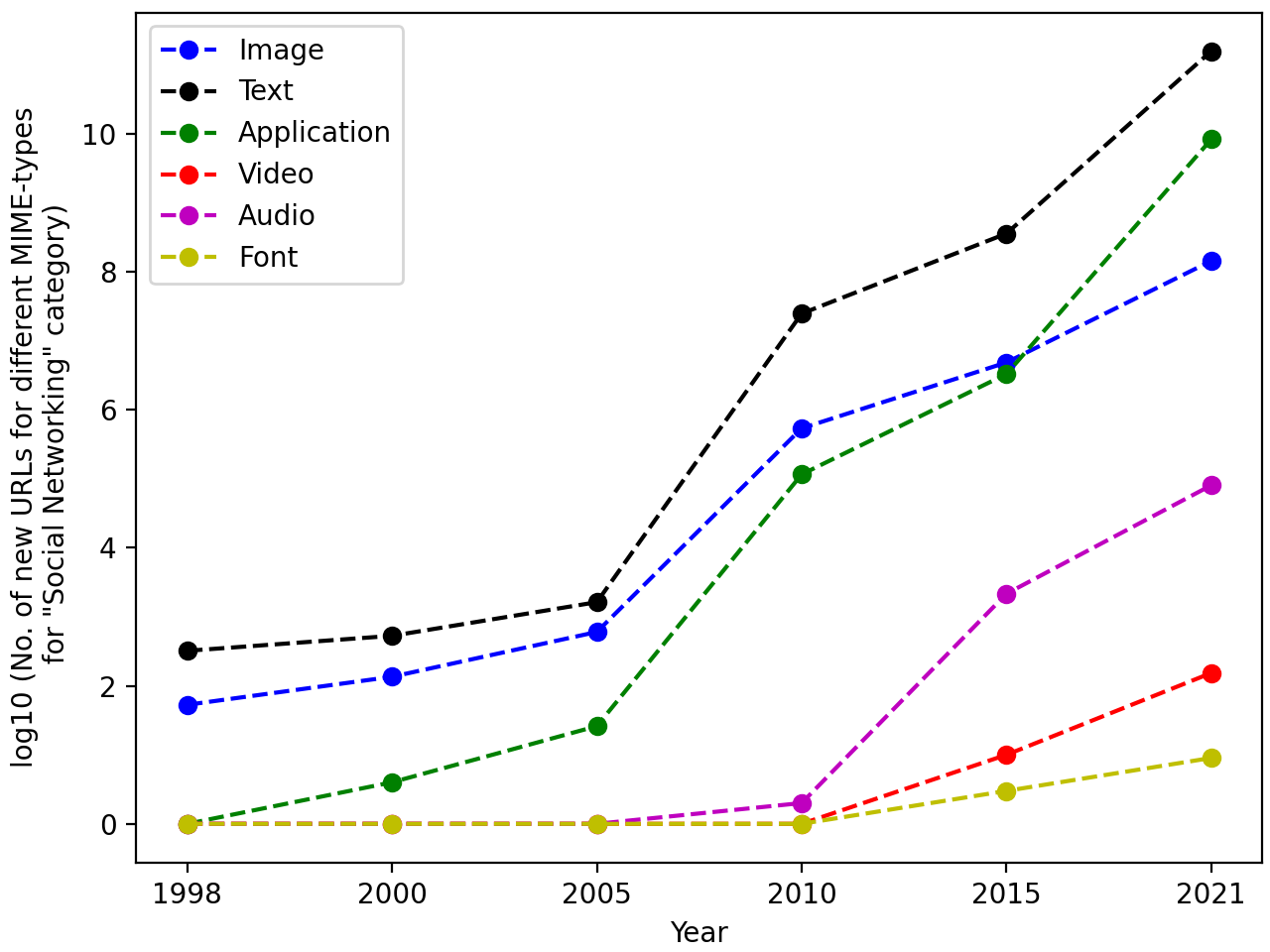}
      \caption{``Social Networking'' category.}
      \label{fig:mime-social-network}
    \end{subfigure}
    \begin{subfigure}{0.45\textwidth}
      \centering
      \includegraphics[width=\linewidth]{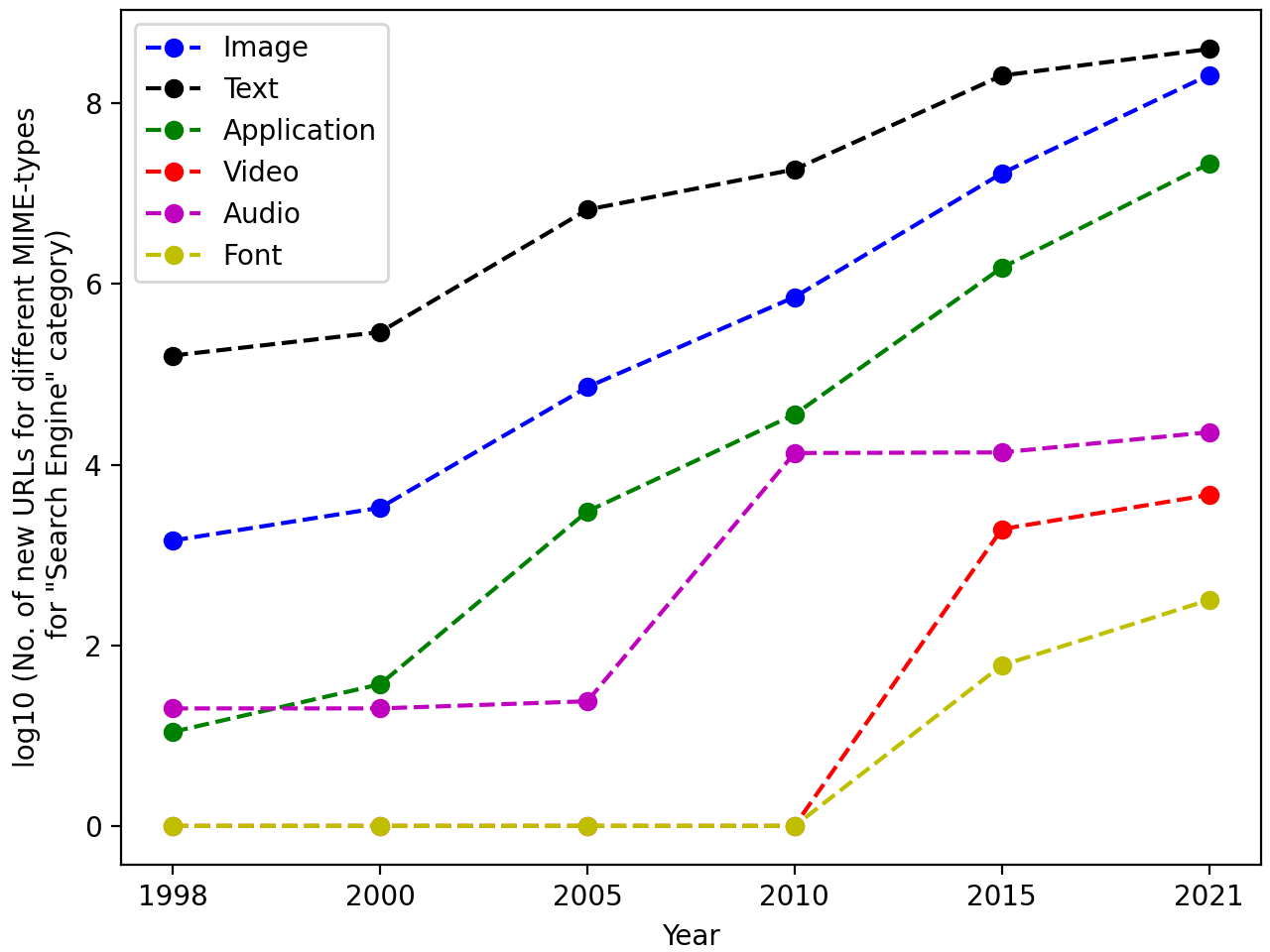}
      \caption{``Search Engine'' category.}
      \label{fig:mime-search-engine}
    \end{subfigure}
\caption{Number of new URLs for different MIME-types for website categories, on log10 scale.}
\label{fig:mime-urls-lineplots}
\end{figure*}

We next turn our attention to the size and complexity of individual websites and look at the number of web resources (URLs) on individual websites, as well as the diversity of the \textit{kinds} of resources (\eg{} text \vs{} image \vs{} video) found at these URLs. \camera{The number of URLs required to fully render a page is indicative of the complexity of the page or site. A rich diversity of MIME types requires more complex code in the browser and ability to display different types of content. The number of MIME types therefore can be seen as another measure of website complexity}.

\begin{figure}{}
      \centering
      \includegraphics[width=\linewidth]{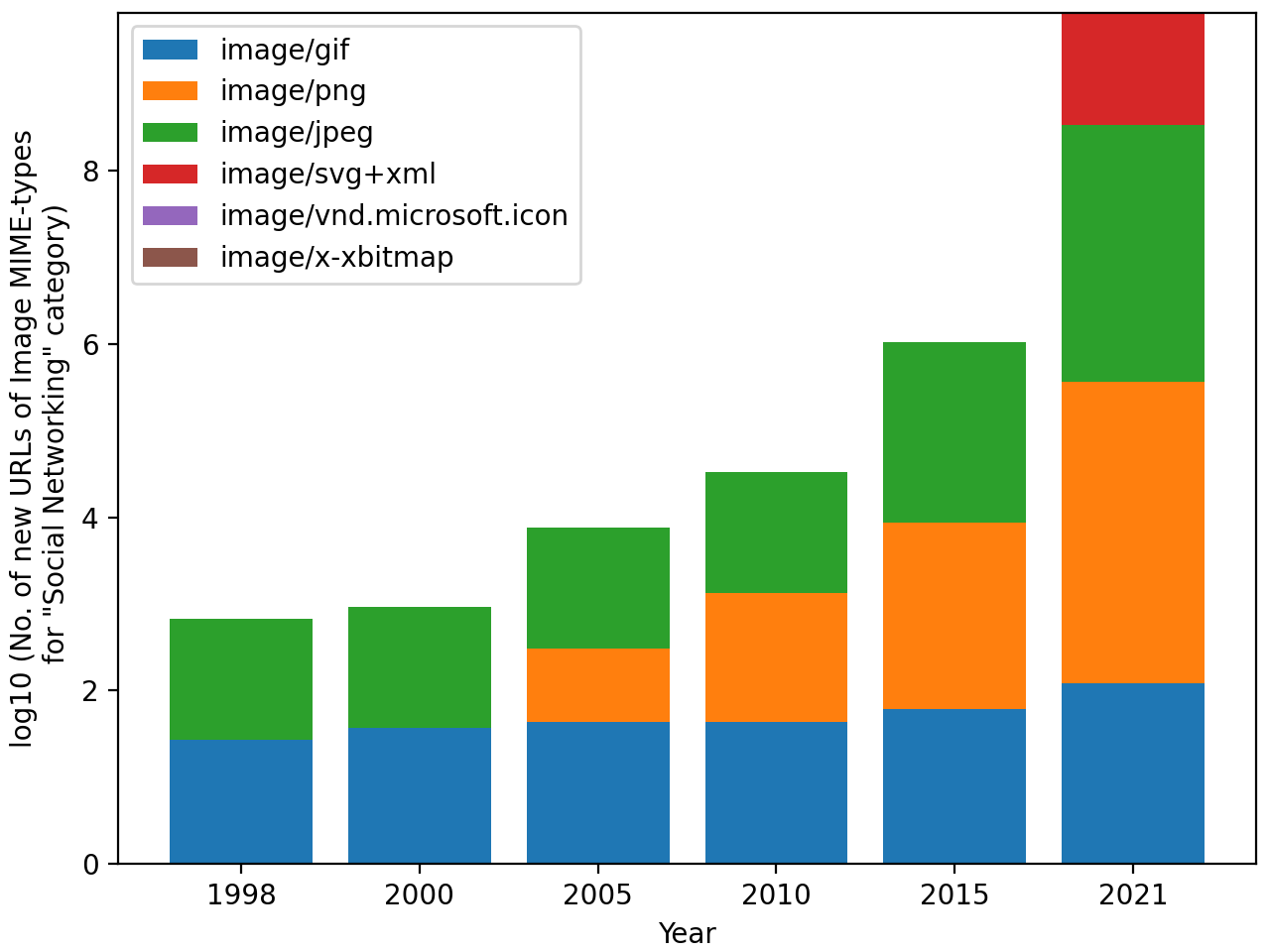}
      \caption{Number of new URLs for Image MIME-type for ``Social Networking'' category.}
      \label{fig:mime-image-sn}
\end{figure}

Figure~\ref{fig:num-mimetypes} captures the overall trend, by looking at the number of \textit{distinct} MIME types represented in URLs of the Alexa Top 100 websites, \camera{collected in Nov 2021}. Since 1998, when about 19 distinct MIME types were present on these top websites, there has been over a five fold increase to 113 different MIME types in 2021.
We also see a rapid increase in the number of different application-specific MIME types. Note that the \textit{number} of distinct MIME types is different from the actual \textit{prevalance} of web resources (URLs) with these MIME types. Figure~\ref{fig:yearly-mime} shows the total raw numbers of \textit{new} URLs (\ie{} URLs added since the previous year) across the Alexa top 100 sites, from 1998 to 2021. This shows a general trend of websites getting increasingly complex with more and more URLs over the years (note that the Y axis is in log-scale). We can also see an increase in usage of web resources with application-specific MIME types, and in addition an increased usage of images over the years. Despite this, it is interesting to see that even in 2021, text-based MIME types (mainly HTML) still constitute a significant proportion of the Web, by number of URLs.  

To further study the usage of images and application-related MIME types, we explore how their usage varies across websites of different categories in Figure~\ref{fig:new-img-urls} and Figure~\ref{fig:new-js-urls} respectively. In both cases, we find that the social networking category has the maximum number of image as well as application (mainly, json and javascript) URLs. Fig~\ref{fig:mime-social-network} digs deeper, showing the numbers of URLs of different MIME types across the years in websites belonging to the social networking category. This clearly shows the rise in URLs of MIME type application/*  starting in 2005, with the rise of AJAX and the RESTful programming paradigm that characterised Web 2.0~\cite{o2009web}. Similarly, the usage and the number of image and other non-text MIME types appears to increase as Internet bandwidth availability increases over the years. We find similar trends for other categories of websites as well. As one example,  Figure~\ref{fig:mime-search-engine} shows that the search engine category follows a similar trend.

Each category of MIME types discussed above (\eg{} image or text) aggregates various different formats of a certain type (\eg{} images can be image/jpeg, image/gif \etc{}). Each such aggregate category itself has an increase in complexity, with new formats being added under each category over the years. Figure~\ref{fig:mime-image-sn} illustrates this for image-related MIME types, by looking at the number of different image-type URLs over the years and the different kinds of image MIME types being used on Alexa top 100 sites, as recorded by \ws{archive.org}. In 1998 and 2000, GIF and JPEG images dominated. Since then, PNG has started to emerge as a common choice of image format. More recently, vector graphics (SVG) have started to be used as well for images. Again, we observe similar trends for other MIME-type classes such as application/* (not shown).

\section{Discussion and Conclusions}
\label{sec:discussion}
 Through the Web, humanity has created a vast treasure trove of data the likes of which has never been seen before. Recently, there have been concerns that such data is being exploited for commercial and other more malicious ends. \camera{For example, detailed online historical profiles of the users are created by various third-parties~\cite{hu2020multi, hu2019websci, hu2020websci}, mainly to target ads and generate revenue~\cite{agarwal2021under, agarwal2020stop, vekaria2021differential}. It also poses various privacy issues for the users online~\cite{beigi2019identifying}}. However, this data revolution has also presented the historians with unparalleled amounts of primary data, which can yield interesting insights and new ways to develop historical understanding of our digital lives.

 To take advantage of this data, we need to develop new ways to glean the historical patterns present in the data.  This work represents a first attempt to develop a quantitative understanding of the evolution of the Web. We looked at several sources of data, ranging from the periodic snapshots of websites on \ws{archive.org} to volumes of searches for different keywords, as captured on Google Trends. To keep this study tractable, we also focused on the most popular websites, using rankings such as Alexa Top websites.

 Our initial study provided quantitative evidence for several interesting patterns, such as the decline in popularity of traditional news media and the increasing usage of social and streaming media instead. We also showed how the Web has become increasingly complex with different kinds of media types being used over the years. Websites can also change in popularity over time and we showed how Google Trends can be a simple yet effective tool to chart this, for example on social networking websites.

 We believe that the techniques we developed in this paper, simple though they are, have potential to be developed further for a more thorough and in-depth understanding of the Web and its effects on humanity and society. These avenues for future work also highlight important limitations of our current work: For example, as a first attempt at a quantitative historical understanding, we only focus on the Alexa Top 100 websites. However, \ws{archive.org} has data about many other websites which we plan to study in future work. Similarly, the common crawl\footnote{\url{https://commoncrawl.org/}, last accessed 26 Jan 2022.} represents a huge dataset of Web URLs and their content, that can provide additional information. Alexa ranking is also not the only and the definitive ranking of websites; thus results may be slightly different if other rankings of Websites (\eg{} Similarweb) are used. \camera{Also, we do not know the demographics of the users or the kinds of devices they use for browsing with Alexa plug-ins that generate Alexa rankings and Websites traffic information}.
 Secondly, our focus on the ``worldwide'' top websites does not, in general, include sites that may be more relevant in non-western cultures, although some populous countries such as China are already represented in the Global top 100 through websites such as Baidu, Weibo and Alibaba. Future work may want to focus on country- or region-specific issues. Our list is also heavily biased towards the English Web, so a study of history of the Web in other languages is an important understudied problem. Other data sets we use, such as Google Trends, also introduce biases into our study: \eg{} Google Trends does not capture the interests of users of search engines other than Google; thus it becomes important to check whether other search engines such as Bing or DuckDuckGo see similar trends as seen on Google. Finally, we also note that a quantitative study can miss several of the nuances that come from close study in the social sciences, thus there is a need to bridge Big Data methods such as ours with more traditional social science approaches~\cite{karamshuk2017bridging}. Qualitative studies and approaches such as the regularly conducted Pew Internet Surveys\footnote{\url{https://www.pewresearch.org/politics/methodology/collecting-survey-data/internet-surveys/}, last accessed 26 Jan 2022.} add important dimensions that cannot be easily captured by purely observational data-based studies such as ours.

\bibliographystyle{ACM-Reference-Format}
\bibliography{main}










\end{document}